\documentclass[12pt,dvips]{article}
\usepackage{amscd,verbatim}
\usepackage[all]{xy}
\usepackage{graphicx}

\usepackage{amsmath}
\usepackage[psamsfonts]{amssymb}
\usepackage{amsthm}
\usepackage{euscript}

\usepackage{latexsym}

\renewcommand{\(}{\begin{equation}}
\renewcommand{\)}{end{equation} \vspace{-.05in}\linebreak}

\newcounter{saveeqn}
\newcounter{savealpheqn}

\newcommand{\alpheqn}{\setcounter{saveeqn}{\value{equation}}%
  \stepcounter{saveeqn}\setcounter{equation}{0}%
  \renewcommand{\theequation}{\mbox{\arabic{section}.\arabic{saveeqn}
\alph{equation}}}
  \renewcommand{\)}{\end{equation}}}
\def\part#1{\frac{\partial}{\partial{#1}}}%
\def\group#1{\refstepcounter{equation}\setcounter{saveeqn}{\value{equati
on}}%
  \label{#1}\setcounter{equation}{0}%
\renewcommand{\theequation}{\mbox{\arabic{section}.\arabic{saveeqn}
\alph{equation}}}
  \renewcommand{\)}{\end{equation}}}
\newcommand{\reseteqn}{\setcounter{equation}{\value{saveeqn}}%
  \renewcommand{\theequation}{\arabic{section}.\arabic{equation}}%
  \renewcommand{\)}{\end{equation}}}

\newcommand{\aalpheqn}{\setcounter{saveeqn}{\value{equation}}%
  \stepcounter{saveeqn}\setcounter{equation}{0}%
  \renewcommand{\theequation}{\mbox{
        \Alph{subsection}.\arabic{saveeqn}\alph{equation}}}
   \renewcommand{\)}{\end{equation}}}
\newcommand{\areseteqn}{\setcounter{equation}{\value{saveeqn}}%
  \renewcommand{\theequation}{\Alph{subsection}.\arabic{equation}}%
  \renewcommand{\)}{\end{equation}}}

\renewcommand{\thefootnote}{\alph{footnote}}
\renewcommand{\(}{\begin{equation}}
\renewcommand{\)}{\end{equation}}
\newcommand{\ba}{\begin{eqnarray}}
\newcommand{\ea}{\end{eqnarray}}

\newcommand{\bp}{\mathop{\vtop{\ialign{##\crcr
   $\hfil\displaystyle{}\hfil$\crcr\noalign{\kern-13pt\nointerlineskip}
   \BIG{(}\hskip0pt\crcr\noalign{\kern3pt}}}}}
\newcommand{\cbp}{\mathop{\vtop{\ialign{##\crcr
   $\hfil\displaystyle{}\hfil$\crcr\noalign{\kern-13pt\nointerlineskip}
   \BIG{)}\hskip0pt\crcr\noalign{\kern3pt}}}}}
\newcommand{\pa}{\mathop{\vtop{\ialign{##\crcr

$\hfil\displaystyle{\oplus}\hfil$\crcr\noalign{\kern+1pt\nointerlineskip
}
   \hspace{.08in}$^{\alpha=0}$\hskip6pt\crcr\noalign{\kern3pt}}}}}

\newcommand{\beq}{\begin{equation}}
\newcommand{\eeq}{\end{equation}}

\numberwithin{equation}{section}

\def\hsp#1{\hspace{#1in}}

\catcode`\@=11
\def\vereq#1#2{\lower3pt\vbox{\baselineskip1.5pt \lineskip1.5pt
\ialign{$\m@th#1\hfill##\hfil$\crcr#2\crcr\sim\crcr}}}
\catcode`\@=12

\makeatletter
\newcommand\figcaption{\def\@captype{figure}\caption}
\newcommand\tabcaption{\def\@captype{table}\caption}
\makeatother
\renewcommand{\(}{\begin{equation}}
\renewcommand{\)}{\end{equation}}

\newcommand{\RR}{{\mathbb R}}
\newcommand{\ZZ}{{\mathbb Z}}
\newcommand{\QQ}{{\mathbb Q}}

\theoremstyle{plain}

\theoremstyle{definition}

\begin{document}

\begin{titlepage}
\begin{flushright}

hep-th/0509046
\end{flushright}

\vspace{2em}
\def\thefootnote{\fnsymbol{footnote}}

\begin{center}
{\Large\bf Duality symmetry and the form fields of M-theory}
\end{center} \vspace{1em}

\begin{center}
{\large Hisham Sati} \footnote{E-mail: {\tt hisham.sati@adelaide.edu.au}\\
Research supported by the Australian Research Council. }
\end{center}

\begin{center}
\vspace{1em} {\em {
Department of Physics\\
and\\
Department of Pure Mathematics\\
       University of Adelaide\\
       Adelaide, SA 5005,\\
       Australia\\
\hsp{.3}\\
Department of Theoretical Physics\\
Research School of Physical Sciences and Engineering\\
The Australian National University\\
Canberra, ACT 0200\\
Australia}}\\

\end{center}

\vspace{0em}
\begin{abstract}
\noindent 

In previous work we derived the topological terms in the
M-theory action in terms of certain characters that we defined. In
this paper, we propose the extention of these characters to include the 
dual fields.
The unified treatment of the M-theory four-form field
strength and its dual leads to several observations. In particular  
we elaborate on the possibility of a twisted cohomology theory 
with a twist given by degrees higher than three.

\end{abstract}

\vfill

\end{titlepage}
\setcounter{footnote}{0}
\renewcommand{\thefootnote}{\arabic{footnote}}

\pagebreak
\renewcommand{\thepage}{\arabic{page}}

\section{Introduction}


Interesting global information is encoded in the Maxwell-like rank four
field $G_4$ of M-theory, which is written locally as $G_4=dC_3$ where
$C_3$ is the so-called C-field. So one concrete aim in this direction is
to understand the nature of this C-field. Another is to understand
Hodge duality that relates $G_4$ to its dual $*G_4$ in eleven dimensions.
There is an analogous question in type II string theory where the fields
are grouped into a total field strength containing the fields descending
from $G_4$, by dimensional reduction, as well as their (ten-dimensional)
Hodge duals. This package leads to the description in terms of twisted
K-theory \cite{MW} \cite{DMW} \cite{MS}.

\vspace{3mm} 
We would like then to ask whether,
in analogy to the type II case, we can unify both field strengths in
eleven dimensions, namely the fields $G_4$ and $*G_4$.  
So we seek a generalized cohomology theory in which the
eleven-dimensional fields are unified in the same way that the
Ramond-Ramond fields (in the presence of Neveu-Schwarz fields) are
unified into (twisted) K-theory.
Earlier work \cite{KS1} \cite{KS2} \cite{KS3} with I. Kriz
viewed elliptic cohomology as the right setting for type II string theory.
The corresponding picture in M-theory leads to the question of whether the
theory $\mathcal M$ proposed in \cite{S1} is new or whether it happens to
be one of  the known generalized cohomology theories. In \cite{S2} we
proposed a unified
quantization condition on $G_4$ and its dual by viewing the
pair
as components of the same total field strength. So the
point we look at in the present paper is the possibility that this
total field strength `lives' in some generalized cohomology
theory.

\vspace{3mm} One might argue that the problem can be looked at 
from the complementary picture of branes. In the same way that one 
has to
talk about branes up to creation of other branes in type II string
theory \cite{MM}, here we ask whether one can talk about M-branes
up to creation of other M-branes. While the picture is not
precisely analogous, one can say that the existence of the
$M5$-brane automatically requires the existence of the $M2$-brane,
via the Hanany-Witten mechanism or via the dielectric effect. \footnote{ This
was discussed briefly in \cite{ES}.}

\vspace{3mm} The supermultiplet $(g_{\mu \nu},\psi_{\mu}, C_3)$ of
eleven-dimensional supergravity \cite{CJS} is composed of the
metric, the gravitino and the C-field. Thus, in its standard
formulation, the theory is manifestly duality-nonsymmetric. One can
then ask about the role of the dual fields in the theory. One can
get a free supersymmetric theory based on the dual 6-index field
$C_6$, but the corresponding interacting theory is not consistent
\cite{NTV}. There is also a duality-symmetric formulation of
eleven-dimensional supergravity \cite{Ban}. \footnote{For
ten-dimensional supergravity theories, this was dicussed in
\cite{C1}, \cite{C2} and \cite{faces}.} However, such a formulation
does not seem to accommodate nontrivial topology or fields that are
nontrivial in cohomology. There is also the duality-symmetric
formulation of the nongravitational fields in \cite{dual}, again
assuming $G_4=dC_3$, i.e. the field $G_4$ is trivial in cohomology,
$[G_4]=0$. 


\vspace{3mm} 
We need a degree
four `Bott generator' and either a degree seven or a degree eight gnerator
for the dual. 
Using the rank seven field $*G_4$ as the dual field, we find the equations
of motion (henceforth EOM) and the Bianchi identity as components of a
unified expression of the total field strength, using a twisted
differential, with the twist now given by the degree four field $G_4$
instead of $H_3$, in the usual case of type II string theory. Adding the
one-loop term $I_8$ to the EOM serves a priori as an obstruction to having
such a twisted cohomology. However, by absorbing $I_8$ in the definition 
of
the dual field strength one still gets a twisting.


\vspace{3mm} One can ask about the relevance of the $E_8$ gauge
theory. We know that the degree four field $G_4$ is intimately
related to $E_8$, at least topologically \cite{W1}. What we are
advocating is that there two ways of looking at the problem, one via
$E_8$, and another via some generalized cohomology theory. But then
adding the dual fields, one seems to break that connection, and in
this case it seems possible to only look for a generalized
cohomology interpretation, as the homotopy type of $E_8$ does not
allow for a direct interpretation of the dual field(s).

\vspace{3mm} So we argue for two points of view regarding the fields.
The first is the bundle picture in which only the lower-rank fields
`electric' fields are described, e.g. $G_4$  in M-theory via $E_8$,
$F_2$ in type IIA via the M-theory $S^1$-bundle. The second is the
generalized
cohomology picture where the field strengths and their duals are grouped
into one total field strength that lives in the corresponding generalized
cohomology theory, e.g. twisted K-theory for type II. Thus taking the second
point of view, the aim of this paper is to argue for a generalized cohomology
theory for the case of the M-theory field strength $G_4$ and its dual.
Such a unification was already started in \cite{S2} where the class of $G_4$
and the dual class $\Theta$ (realizing the RHS of the EOM) were given
a unified expression that reflected their quantization laws. The existence of the
corresponding generalized cohomology theory was proposed in \cite{S1} and
further properties were given in \cite{S2}.

\section{The total field strength}

First note that, unlike the RR fields which have mod 2 periodicity,
the fields of M-theory do not enjoy such a periodicity. This is
obvious because one of the fields has even rank and the other has
odd rank. Besides there are only two of them. One can ask first
whether there is a Bott element of dimension three ($=$ the
difference of the two ranks) that can take the role which the usual
Bott element played in type II. The answer is negative and there is
no such element in the class of theories descending directly from
$MU$. So one can then ask whether there is another way to form a
total M-theory field strength with a uniform degree. One is then
forced to use more than one element to do the job. Again there is no
element of odd degree, so in order to be able to say something
useful, one seeks a modification of the point of view in which even
degree fields are included. But what exactly should we do? Two
things come to mind. First we can try to lift to the bounding twelve
dimensional theory defined on $Z^{12}$ with $\partial
Z^{12}=Y^{11}$. Here, one possibility is then to look at the
four/eight combination $G^{(12)}=G_4 + *_{12}G_4$ in twelve
dimensions. Then the arguments that hold for $G_4 + d*_{11}G_4$ in
eleven dimensions hold for $G^{(12)}$ as well. \footnote{Throughout
the paper, if the Hodge star operator has no explicit dimension
label then it refers to the eleven-dimensional one.} Second, we can
work with an eight-form in eleven dimensions, that we view as the
dual field instead of the seven form. On the other hand, if we
insist on working with odd forms, then this seems to suggest some
deformation of cohomology rings which involves odd generators.

\vspace{3mm}
We are looking for a generator of degree four that makes a degree
zero form when multiplied with $G_4$. 
Since ${\rm dim}v_n=2p^n-2$, there is only one generator of degree four,
which the first generator at $p=3$. What theory is a good candidate
theory to include this generator? It is possible that this is  
either of the
first Morava K-theories at $p=3$, i.e. either ${\widetilde K}(1)$ or
$K(1)$ with coefficient rings ${\widetilde
K}(1)_*=\ZZ[v_1,v_1^{-1}]$, and ${K}(1)_*=\ZZ/3[v_1,v_1^{-1}]$,
respectively. We can then form the desired class \footnote{In
writing this expression and all the analogous ones, we are
implicitly tensoring with $\RR$ (or $\QQ$).}
 \( (v_{1,p=3})^{-1}G_4.\)

\vspace{3mm}
As in the case for $G_4$ we are looking for a generator whose
degree is the same as the degree of the field, and which is inverted
so that its inverse can be used to write down a uniform degree
zero field. So here we need a degree eight generator. 
Now we would like to find an expression of total degree zero for the
total M-theory field strength. 
The desired generator is the square of
$v_{1,p=3}$, which has total dimension $4+4=8$. So with this
possibility, we can write the following expression for
the uniform total field strength 
\footnote{This is meant to be analogous to the uniform degree zero
expressions of the RR field strengths in \cite{F}.}
\( G=(v_{1,p=3})^{-1}G_4 +
(v_{1,p=3})^{-2}G_8. \) With this, we are using the same generator
for the whole expression, which is the case analogous to the type II
situation, 
One possibility that that we are then
dealing with the $p=3$ first (integral) Morava K-theory. 
One can ask whether the problem can be looked at
without specializing to a particular prime. The theory of
Topological Modular Forms, $tmf$, 
has an interesting feature that it is not
localized at a given prime, i.e. is not local and unifies all primes
-- see \cite{KS3} for a discussion on the relevance of TMF from a
different but related point of view. This is attractive, and seems
to be what a theory like M-theory should be doing. Besides, this
might make sense since the vector bundles (or their `higher-degree'
analogs) are real, and TMF is a real theory -- it is to elliptic
cohomology $E$ as $KO$-theory is to $K$-theory.
One can ask whether there are degree four and degree
eight generators in $tmf$, which can be used for the total field
strength. 
Indeed there are such generators, which were used in \cite{KS3}.

\vspace{3mm}
As far as dynamics goes, it does not make much sense
to talk about $*G_4$ or $d*G_4$ alone, because their dynamics involve
$G_4$ (cf. the EOM of $G_4$). So in order to include the dual picture,
one can at best look for a duality-symmetric formulation of the
character,
i.e. as opposed to a dual description.
If we use the eight-form $d*G_4$ as the `dual' form, then the
corresponding exponential is \( e^{G_4 +d*G_4}. \label{8}\) We ask
the question whether from this we can get the EOM and the Bianchi
identity. By looking at the degrees of the forms, we see that while
we can get the Bianchi identity by looking at the degree five
component, i.e. \( \left[d\left(  e^{G_4 +d*G_4} \right)
\right]_{(5)}, \) we cannot get the EOM, simply because the
degrees of forms would not match. \footnote{unless the
differential does not act on the exponential, which is not what is
meant to happen.}

\vspace{3mm}
One can then ask whether the exponential (\ref{8})
can be looked at in some other way that would give the EOM and
Bianchi. While the EOM can be obtained by some `flatness condition' on
the character, i.e. \( \left[e^{G_4 + d*G_4} \right]_{(8)}=0,
\label{flat}\) the Bianchi identity does not follow.
One instead gets a flatness condition on $G_4$ as well if one were
to look at the degree four component of the expression (\ref{flat}).
Even though one can say we got both the EOM and the
Bianchi identity, we actually did not do that by using the same
expression, and this is obviously not satisfactory. This seems to
indicate
that while the quantization conditions on the forms \cite{S2} favors the
four/eight combination, the dynamics favors instead the four/seven
combinations of field strengths.

\vspace{3mm} Let us now look at the effect of including the
generators-- let us call them $v$ and ${\widetilde v}$ -- in
(\ref{8}). Doing so results in the expression \( \left[\left(
e^{v^{-1}G_4 + {\widetilde v}^{-1}d*G_4} \right)
\right]_{(8)}=\frac{1}{2}v^{-2} G_4 \wedge G_4 + {\widetilde
v}^{-1} d*G_4. \label{vv} \) So requiring that we get the EOM via
factoring out the generators leads to the obvious condition that
\footnote{It is interesting that if we interpret $v$ and 
${\widetilde v}$ as the generators introduced in \cite{dual} and
used in the next section, then the corresponding statement 
would be $\{v,v\}=-{\widetilde v}$, i.e. one of the relations 
of the gauge algebra for $G_4$ and $*G_4$. The minus sign would then
make (\ref{vv}) equal to $(d+ \mathcal{G})^2$, the obstruction to 
nilpotency.} 
\( v^2={\widetilde v}. \) Naturally, we would like to see whether
such a condition can occur in the generalized cohomology theories
that we consider in this paper. We check the dimensions of the
generators. Since in general that dimension at `level' $n$ and
prime $p$ is ${\rm dim}~v_n=2(p^n-1)$, we then need to
satisfy the equality \( \left[2(p^n-1)\right]^2=2(p^m-1),
\label{sq}\) where $m>n$. Even though such an expression is not
expected to have many solutions in general, it is still more
general than we want.

\vspace{3mm} It might be desirable to require that the total
expression on the RHS of (\ref{vv}) have degree zero. It turns out
that this is not possible within the current context, and the next
best thing is to require the first generator $v$ to have degree
four. \footnote{In any case, even without requiring the $G_4$ term
to have degree zero, one sees upon inspecting (\ref{sq}), at least
for relatively low $n$, $m$ and $p$ (which are the only relevant),
that the result of the discussion does not change.}
 This then implies,
via $2(p^n-1)=4$, that $p=3$ and $n=1$. Of course the equality is
then satisfied and the dimension of ${\widetilde v}$ is $16$ with
$m=2$ and the same prime $p=3$.

\vspace{3mm} Let us go back and look at what the above implies for
the relationship between the dimensions of the generators and the
dimensions of the field strengths. In the above we asked
whether the expanded exponential expression has total degree zero.
But then going back to the exponent, we see that it does not have
total degree zero, because we have the generator ${\widetilde v}$,
which we found to have dimension sixteen, multiplying $d*G_4$
which has rank eight as a form or a class. However, it is still
true that the $G_4$ part has degree zero. What we learn from this
is that what matters is for the degrees of the factors to match
after expanding the exponential and not as they stand in the
exponent. As mentioned earlier, generators of degree four and eight
can be obtained from tmf (cf. \cite{KS3}).

\section{A twisted (generalized) cohomology?}
\label{deg7}

In this section, we would like to use the degree seven field as
the dual field to $G_4$ and thus take the total field strength to
be $G=G_4 + *_{11}G_4$. We would like to use such an expression (and 
slight variations on it --see below) as it is duality-symmetric
\footnote{This is meant to be in the sense that the expression contains
both $G_4$ and its dual $*G_4$, and that it is invariant under the 
exchange $G_4 \leftrightarrow *G_4$. It is not meant to be in the 
sense of exchanging $G$ and $*G$ as we will see explicitly later when 
the generators of the gauge algebra are included.}
in the electric-magnetic or membrane-fivebrane sense.
Then it is interesting that one can write
the Bianchi identity and the EOM of $G_4$, respectively, as the degree
five and the degree eight component
of the expression \( \left( d+ \frac{1}{2} G_4\wedge 
+\frac{1}{2}*G_4\wedge \right) G=0.
\label{twist}\) The degree eleven component, i.e. the cross-terms 
between $G_4$ and $*G_4$, vanish because of the relative minus sign,
\footnote{since $\alpha_k \wedge \beta_l=(-)^{kl}\beta_l \wedge 
\alpha_k$.} and the $*G_4\wedge *G_4$ term vanishes because it involves
the same form of odd degree. 

\vspace{3mm}
There are several interesting aspects to equation (\ref{twist}). First, 
one can ask whether this has the form of
some twisted structure in analogy to that associated with the RR
fields in type II string theory, where one has for the total field
strength $F$, \( dF=H_3 \wedge F.\) Written as \( d_{H_3}F=(d-H_3
\wedge)F=0,\) this leads to interpreting $d-H_3$ as the
differential in twisted (de Rham) cohomology $H^*(X,H;\RR)$, even
for type IIA and odd for type IIB \cite{MS}. One can easily check
that $(d_{H_3})^2$ is indeed zero \cite{MS}, which follows
from the fact that the twisting field $H_3$ is closed and that the wedge
product of two twisting fields $H_3\wedge H_3$ vanishes just
because it is the wedge product of the same differential
form of odd degree.

\vspace{3mm} Going back to (\ref{twist}), we ask whether an
analogous structure appears. Of course we have obvious differences
from the type II case: what is to be interpreted as a `twisting
field', $\frac{1}{2}G$, is now part of the total field that is being
twisted, namely $G$.
\footnote{In order to make the equations and the statement symmetric, 
one might try to rescale and use both the total field strength and the
twist as $\frac{1}{\sqrt{2}}G$. However, the equation of motion 
would then have an anomalous relative factor of $\sqrt{2}$.} 
 The other difference is that the twist now involves an
even rank field, which while it is closed in analogy to $H_3$, the
wedge of two copies of which does not vanish since it is
even-dimensional. If we interpret the combination $d+
\frac{1}{2}G_4+\frac{1}{2}*G_4$ as a new differential $d_{G}$ and hope 
that it forms a cohomology, then the nilpotency does not seem to be
immediately obvious. However, it turns out that the situation is in fact
encouraging. To see this, let us simply calculate the
action of its square on the total field strength,
\(
d_{G}^2~G=\left( d+ \frac{1}{2}G_4\wedge +\frac{1}{2}*G_4 
\wedge \right)^2 G, 
\)
but this is zero for the same reasons that equation (\ref{twist}) holds,
namely by use of the EOM and the Bianchi identity, and by the fact that
the rest of the terms have high degrees. Thus,
\(
d_{G}^2~G=0.\)
This is on-shell and is valid 
when the differential acts on the field strength. In the case of type II
string theory, $d_H=d+H_3$ was an actual differential, i.e. $d_H^2$ was 
zero without necessarily acting on the RR field $F$. Does this happen in 
our case of M-theory? 

\vspace{3mm}
Let us study the question one step at a time. To start, calculating
$d_G^2$ gives the sum
\(
\frac{1}{2}G_4\wedge d + \frac{1}{2}G_4 \wedge d,
\)
i.e. $G_4\wedge d$. Obviously this is not zero, and so we need 
to modify the differential in order to have any hope at nilpotency. 
The problem can be traced back to the fact that $G_4$ has an even degree
and so moving the differetnial over it does not pick a minus sign that 
would then cancel the other factor. Explicitly, the square gives the 
cross terms $d(G_4\wedge) + G_4\wedge d$, which when expanded gives
$dG_4\wedge + G_4 \wedge d + G_4\wedge d$. The first term disappears 
because of the Bianchi identity but the second {\it adds} to the third
(instead of {\it subtracting} had $G_4$ been of odd degree).
Thus the problem does not arise for $*G_4$.
Note that at this stage we can see that the somewhat artificial factor
of half inside the differential does not seem to matter. We will see
that this is indeed the case later.  

\vspace{3mm} In order to get the two terms above to subtract instead of 
add, we need some form of grading. For that purpose, let us use the 
duality-symmetric total field strength introduced in \cite{dual},
\(
\mathcal{G}=vG_4+{\widetilde v} *G_4, 
\)
and check whether this $\mathcal{G}$ can be used as a twist to form the 
desired differetial. As the problem above was due to the sign in the 
Leibnitz rule, let us consider the corresponding rule for $\mathcal{G}$.
Due to the nature of $v$ and ${\widetilde{v}}$ \cite{dual}, this is
\(
d(\mathcal{G}\wedge)=d\mathcal{G} \wedge ~ -\mathcal{G} \wedge d. 
\)
Then using this Leibnitz rule to expand the expression
\(
(d\pm \mathcal{G})^2=d^2 \pm d (\mathcal{G}\wedge) \pm 
\mathcal{G}\wedge d + \mathcal{G}\wedge \mathcal{G}
\)
gives
\(
(d\pm \mathcal{G})^2=\pm d\mathcal{G} + \mathcal{G}\wedge \mathcal{G}.
\)
Now which sign to pick is determined simply by the vanishing of the
right hand side. This happens for the minus sign 
\footnote{One way this minus sign can be motivated is by saying it 
gives the differential $d- H$ in type IIA upon reduction (at least of
the $G_4$-part of $\mathcal{G}$.}
because then the right hand side would be 
\(
d\mathcal{G} - \mathcal{G} \wedge \mathcal{G},
\)
which is zero as it is just the negative of the unified equation giving 
the EOM and the Bianchi identity derived in \cite{dual}. Then, 
$d-\mathcal{G}$ is indeed 
a differential, which we will denote by $d_{\mathcal{G}}$. 
At this point we can try to look for slight variations of this 
differential.
\begin{itemize}
\item {\it Scaling}: From the expression 
\(
(d+n\mathcal{G})^2=-nd\mathcal{G} + n^2 \mathcal{G}\wedge{G}
\)
we see that the constant $n$ can only be equal to one in order for the
unified equation of motion to be satisfied.
\item {\it Duality}: We can derive the Leibnitz rule for the dual field
$*\mathcal{G}$,
\(
d(*\mathcal{G}\wedge)=d(*\mathcal{G})\wedge + *\mathcal{G}\wedge d,
\)
which we use to show that 
\(
(d\pm n* \mathcal{G})^2=\pm n d*\mathcal{G}\wedge~ 
\pm n *\mathcal{G}\wedge d ~\pm n *\mathcal{G}\wedge d~
+ n^2 *\mathcal{G} \wedge *\mathcal{G}.
\label{star}
\)
It is obvious then that $(d \pm n* \mathcal{G})$ is not a differetial 
since the terms $\pm n *\mathcal{G}\wedge d$ in (\ref{star}) add, giving
a result that cannot be zero without acting in a particular way on 
other forms.
\end{itemize}

\vspace{3mm} So does this mean we have twisted cohomology? This
suggests that one gets such a structure if one uses the
rank seven field $*_{11}G_4$ as the dual field of the M-theory
rank four field $G_4$. At the level of differential forms, the differential
$d_{{G}}$ is then interpreted as a map
\(
d_{{G}}: \Omega^m \oplus \Omega^{m-3} \longrightarrow \Omega^{m+1}
\oplus \Omega^{m-2},
\)
our case being $m=7$ of course. Such differentials
(with one twist) were encountered in \cite{BHM}. One can also form a 
differential of
uniform degree by introducing a formal parameter $t$ of degree $-3$ and
write 
\footnote{We are oversimplifying as we also have to include $v$ and 
$\widetilde{v}$. We hope to discuss this elsewhere.}
$d_{G_4}=d+ t G_4 + t^2 *G_4$. The 
interpretation of $t$ as a 
periodicity generator is desirable but is not very transparent again 
because it is of odd degree. This shift from even to odd degrees can be 
obtained by suspension or by looping (see below for relevance). 

\vspace{3mm} Furthermore, we would like to interpret the above
result at the level of twisted cohomology as the target via a
generalized Chern character of some {\it twisted generalized
cohomology theory} ${\mathcal M}(\bullet,G)$ or
${}^{\tau}{\mathcal M}$, where $\tau$ is $[G]$, the class of
$G$, i.e. \( ch_{G}: {\mathcal M}(\bullet,G) \longrightarrow
H^{4k}(\bullet,G). \)

\vspace{3mm}
Note that elliptic cohomology theory can be thought of, at least
heuristically, as the K-theory on the loop space, i.e.
the elliptic cohomology of a space $X$ is the K-theory of $LX$.
The twists of K-theory are given by its automorphism. This includes
$H^3(X;\ZZ)$. Applying this to the loop space gives the automorphism 
of elliptic cohomology, by which one can twist.
\footnote{We thank Constantin Teleman for explanations concerning this 
point.} By transgression,
$H^3(LX;\ZZ)$ gives $H^4(X;\ZZ)$. For the $d*G_4$ part, 
we expect the arguement to be analogous.
The $H^8$-twist in M-theory would descend to $H^7$-twist in string 
theory.
\footnote{We hope to discuss this in detail elsewhere.}  

\subsection{Including the one-loop term}
\label{one}

The EOM after including the one-loop term (first introduced in
\cite{DLM}) is modified to \( d*G_4=-\frac{1}{2}G_4 \wedge G_4 +
I_8, \) where $I_8=-\frac{p_2 -(p_1/2)^2}{48}$ is the purely
gravitational term, a polynomial in the Pontrjagin classes of the
tangent bundle of the eleven dimensional spacetime $Y^{11}$.

\vspace{3mm}
We can still group together $G_4$ and its dual in the presence of $I_8$. For the degree
four/eight combination we simply add $I_8$ to $d*G_4$  and we are dealing with precisely the
$\Theta$-class studied in \cite{DFM} and \cite{S2}. For the case of the degree four/seven
combination, we can use the fact that $I_8=dX_7$ where $X_7$ is the transgression polynomial
for $I_8$ in degree seven, and write the expressions using $*G_4 + X_7$. For example,
\(
\left(d + \frac{1}{2}G_4  \right)\left[G_4 +(*G_4+ X_7) \right],
\)
with the degree five and degree eight pieces giving respectively the Bianchi identity
and the EOM upon using $dX_7=I_8$. Other formulae follow as well. We can see
that when we add $I_8$ to the picture, it serves as an obstruction to
having a twisted theory. However, if we absorb it in the defintion of the
dual field as above, then we would still get a twist.

\newpage
{\bf Acknowledgements}

 \vspace{3mm}
 The author thanks Igor Kriz for helpful explanations 
on generalized cohomology theories, 
Edward Witten for very interesting discussions, Varghese Mathai for very
useful comments, and Arthur Greenspoon for suggestions in improving the 
presentation. He also thanks the 
organizers of the Oberwolfach
mini-workshop on Gerbes, Twisted K-theory and Conformal Field
Theory, the Simons Workshop in Mathematics and Physics at Stony
Brook, and the Theory Division at CERN for their hospitality during
the final stages of this project. 


\end{document}